\documentclass[a4paper, 12pt]{iopart}
\usepackage{graphicx}
\usepackage{cite}
\usepackage[T1]{fontenc}
\usepackage{multirow}
\usepackage{iopams}

\begin{document}

\title[Voter model on networks with two cliques]
{Voter model on networks partitioned into two cliques of arbitrary
  sizes}

\author{Michael T. Gastner and Kota Ishida}

\address{Yale-NUS College, Division of Science, 16 College Avenue
  West, \#01-220 Singapore 138527}
\ead{\mailto{michael.gastner@yale-nus.edu.sg},
  \mailto{kota\_ishida@u.yale-nus.edu.sg}}

\begin{abstract}
  The voter model is an archetypal stochastic process that represents
  opinion dynamics.
  In each update, one agent is chosen uniformly at random.
  The selected agent then copies the current opinion of a randomly
  selected neighbour.
  We investigate the voter model on a network with an exogenous community
  structure: two cliques (i.e.\ complete subgraphs) randomly linked by
  $X$ interclique edges.
  We show that, counterintuitively, the mean consensus time is
  typically not a monotonically decreasing function of $X$.
  Cliques of fixed proportions with opposite initial opinions
  reach a consensus, on average, most quickly if $X$ scales as
  $N^{3/2}$, where $N$ is the number of agents in the network.
  Hence, to accelerate a consensus between cliques, agents should
  connect to more members in the other clique as $N$ increases but
  not to the extent that cliques lose their identity as distinct
  communities.
  We support our numerical results with an equation-based analysis.
  By interpolating between two asymptotic heterogeneous mean-field
  approximations, we obtain an equation for the mean consensus time
  that is in excellent agreement with simulations for all values
  of $X$.
\end{abstract}
\noindent{\it Keywords\/}: voter model, community structure, consensus
time, heterogeneous mean-field approximation, complex networks

\maketitle

\section{Introduction}
\suppressfloats

\begin{figure}
  \begin{center}
    \includegraphics[width=0.4\textwidth]{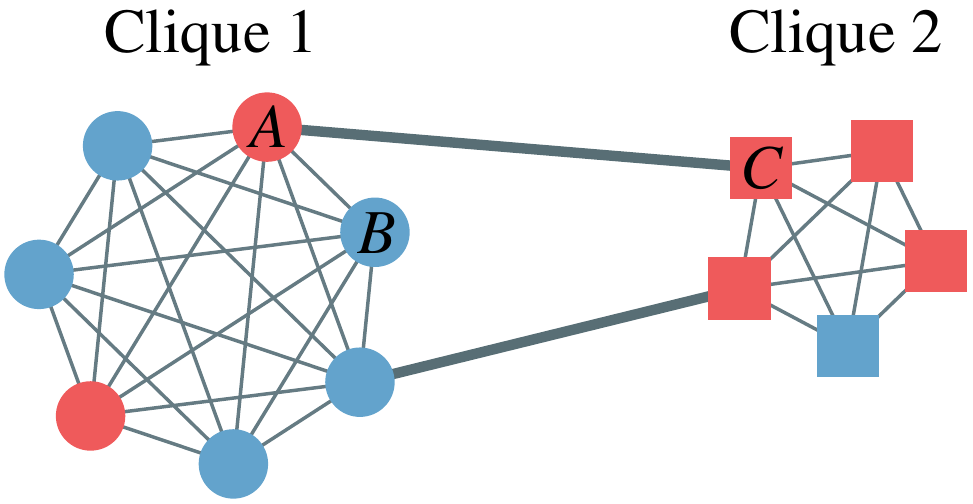}
    \caption{Illustrative example of a small two-clique network.
      In this example, clique 1 is a complete subgraph with seven
      vertices (circles), whereas clique 2 has only five vertices
      (squares).
      Each vertex represents an agent that has exactly one of two
      possible opinions: `red' or `blue'.
      We apply the update rules of the voter model.
      That is, we first choose a random focal vertex, e.g.\ $A$,
      in the depicted network.
      Then, we choose a random neighbour of the focal vertex and copy
      the neighbour's opinion.
      In our example, if the chosen neighbour is $B$, $A$ changes its
      opinion to blue.
      If the chosen neighbour is $C$, $A$ keeps its current (i.e.\
      red) opinion.
      We distinguish between {\em intra}clique edges (thin lines) and
      {\em inter}clique edges (thick lines).
      In our analysis, we vary the relative sizes of the two
      cliques and the number of interclique edges.
    }
    \label{fig:illustrative_example}
  \end{center}
\end{figure}

Opinion formation in social networks has become an active field of
research in statistical physics (for reviews,
see~\cite{castellano_statistical_2009-1, sirbu_opinion_2017,
  jedrzejewski_statistical_2019}).
In particular, the voter model~\cite{holley_ergodic_1975-1,
  redner_reality-inspired_2019} has become a
paradigmatic model for opinion dynamics.
Its rules are simple but sufficiently powerful to reproduce the
summary statistics of real elections~\cite{fernandez-gracia_is_2014}.
In the basic version of the voter model, the vertices on a network
represent agents that can hold exactly one of two possible opinions:
`red' or `blue'.
Repeatedly, one agent is selected independently and uniformly at
random.
This agent then adopts the opinion of a randomly chosen adjacent
agent (\fref{fig:illustrative_example}).
As long as the network is connected and finite, this update
rule guarantees that agents must eventually reach a
consensus~\cite{serrano_conservation_2009}, defined as a state in
which all agents have identical opinions.

The mean time until consensus depends on the initial distribution of
opinions and the network structure.
While early studies of the voter model focused on complete
graphs~\cite{moran_random_1958} or regular
lattices~\cite{clifford_model_1973-1}, interest has recently
shifted towards networks with more complex topologies, e.g.\
small-world networks~\cite{castellano_incomplete_2003,
  vilone_solution_2004}, graphs with right-skewed degree
distributions~\cite{sood_voter_2005, suchecki_voter_2005,
  yang_effects_2009, baronchelli_voter_2011}, multiplex
networks~\cite{diakonova_irreducibility_2016}, or networks with a
community structure~\cite{castello_anomalous_2007, masuda_voter_2014,
  bhat_opinion_2019}.

A subgraph of a network is called a community if there are
significantly more edges within the subgraph and fewer links to the
rest of the network than those predicted by a null model that has no
planted community structure (e.g.\ an Erd\H{o}s--R\'enyi graph with
the same total number of edges as the network under
investigation)~\cite{newman_finding_2004}.
The detection of communities from network data has become a major line
of research with a plethora of different algorithmic
approaches~\cite{fortunato_community_2016}.
Various techniques have confirmed that essentially all networks of
practical relevance contain more than a single
community~\cite{lancichinetti_finding_2011, krzakala_spectral_2013,
  le_estimating_2015, sarkar_eigenvector_2016, riolo_efficient_2017}.

This finding has motivated us to analyse the voter model for one of
the simplest possible types of multi-community networks---namely,
networks with exactly two communities.
Situations where agents divide into two communities are plentiful in
the real world.
For example, a split between communities may arise because of a
language barrier (e.g.\ between Dutch and French speakers in
Belgium~\cite{expert_uncovering_2011}) or differences in race,
ethnicity, age, religion, education, occupation, or
gender~\cite{mcpherson_birds_2001}.
Conversely, agents with similar attributes tend to form close-knit
communities because of status
homophily~\cite{lazarsfeld_friendship_1954}, a social
phenomenon that causes the proverbial birds of a feather to flock
together.
Within communities, cohesion often reaches such an extent that `we
can observe in many groups a social unity within which people feel at
one though their opinions still differ' (p.~229
in~\cite{maciver_society_1949}).

Previously, it has been claimed that the voter model is insensitive to
changes in the community structure~\cite{artime_dynamics_2017}.
This conclusion has mainly rested on results for two equally large
cliques (i.e.\ complete subgraphs), where the mean consensus time is
proportional to the total number of vertices in the network, $N$,
unless the connections between cliques are extremely
sparse~\cite{sood_voter_2008-1, masuda_voter_2014}.
Here, we argue that an investigation of only the special case of equally
large cliques does not do justice to the actual complexity of the
problem.

In this article, we revisit the two-clique voter model but allow
unequal clique sizes.
The dynamics exhibit an intriguing feature: the mean consensus time
is minimal at an intermediate interclique connectivity.
We investigate in detail the case where cliques with given relative
sizes start from opposite opinions, representing a completely polarised
society.
To minimise the mean consensus time, we find that the optimal number
of interclique edges, $X_{\rm min}$, should scale in proportion to
$N^{3/2}$.
This scaling law puts $X_{\rm min}$ between the case of a constant
number of interclique links per agent ($X \propto N$) and a complete
graph ($X \propto N^2$).

After specifying the details of our model in~\sref{sec:model}, we
present the results from numerical simulations
in~\sref{sec:simulation}.
In~\sref{sec:equation-based}, we derive an analytical expression
for the mean consensus time as a function of $X$ for arbitrary clique
sizes.
Our derivation demonstrates how we can go beyond previous
approximations~\cite{masuda_voter_2014, constable_population_2014} to
obtain not only the asymptotic behaviour for either extremely sparsely
or densely connected cliques, but also reliable predictions for
the intermediate interclique connectivity.
We conclude with a discussion of our results in~\sref{sec:discussion}.

\section{Model}
\label{sec:model}

We consider a simple undirected graph with $N$ vertices that
can be partitioned into two cliques, as shown
in~\fref{fig:illustrative_example}.
We denote the fraction of vertices in the first clique by $\alpha \in
(0, 1)$.
The two cliques are connected by $X$ edges randomly selected from
all $\alpha (1-\alpha) N^2$ possible pairs that can be formed by one
vertex in clique $1$ and another vertex in clique $2$.

Each vertex is either red or blue depending on the current opinion of
the corresponding agent.
The time intervals between consecutive opinion updates are
independent, identically distributed exponential random numbers so
that the dynamics are a continuous-time Markov chain.
We choose the time unit such that every individual agent is active
with a rate equal to $1$.

If we wish to keep track of all individual opinions, the cardinality
of the model's state space is $2^N$.
The Monte Carlo algorithm behind all numerical data presented in this
paper is in fact based on this exact agent-based paradigm.
However, summarising and modelling the results at such a fine level of
resolution is neither insightful nor practical given that the number
of configurations grows exponentially with $N$.
Instead, we combine all configurations whose fraction of red agents is
$\rho_1$ in clique $1$ and $\rho_2$ in clique $2$ into the macrostate
$(\rho_1, \rho_2)$ to simplify the data analysis, visualisation, and
mathematical modelling.
Strictly speaking, the Markov chain is not lumpable at this
macroscopic level~\cite{banisch_markov_2015} because we neglect the fact that
different vertices in a clique can be adjacent to a different number
of vertices in the other clique.
Let us denote the number of interclique edges incident on a vertex $v$
by $k_{{\rm inter}, v}$.
The probability distribution of $k_{{\rm inter}, v}$ is approximately
binomial; thus, it is so concentrated near its peak that we
withhold little information if we replace the exact value
$k_{{\rm inter}, v}$ with its mean: $k_{{\rm inter}, v} \approx
X/(\alpha N)$ for all vertices $v$ in clique $1$ and $k_{{\rm inter},
  v} \approx X/[(1-\alpha) N]$ for every $v$ in clique $2$.
%
In the parlance of statistical physics, we apply a heterogeneous
mean-field approximation~\cite{serrano_conservation_2009,
  moretti_heterogenous_2012}: we correctly account for the difference
in the clique sizes and replace the exact microscopic interactions
with an average over the cliques.
More elaborate approximations are
conceivable~\cite{masuda_voter_2014}, but the mean-field approximation
is remarkably accurate, as we will see shortly.
On balance, we do not find the notational burden of a more detailed
approximation to be worth the effort.

\begin{table}
  \caption{\label{tab:Q}Transitions from the state $(\rho_1, \rho_2)$
    and their rates.}
  \setlength{\tabcolsep}{9pt}
  \begin{tabular*}{\textwidth}{@{}lll}
    \br
    New state
    & How is the new
    & Transition rate matrix element\\
    $(y, z)$
    & state reached?
    & $Q[(\rho_1, \rho_2), (y, z)]$\\
    \mr
    \multirow{2}{*}
    {$\left(\rho_1 + \frac{1}{\alpha N}, \rho_2\right)$}
    & A blue agent in clique $1$ adopts
    & \multirow{2}{*}{$\alpha N (1-\rho_1) \cdot
      \frac{\alpha N (\alpha N - 1) \rho_1 \, + \, X \rho_2}
      {\alpha N (\alpha N - 1) \, + \, X}$}\\
    & the opinion of a red agent.\\[6pt]
    \hline
    \ms
    \multirow{2}{*}
    {$\left(\rho_1 - \frac{1}{\alpha N}, \rho_2\right)$}
    & A red agent in clique $1$ adopts
    & \multirow{2}{*}{$\alpha N \rho_1 \cdot
      \frac{\alpha N (\alpha N - 1) (1 - \rho_1) \, + \, X (1 -
      \rho_2)}
      {\alpha N (\alpha N - 1) \, + \, X}$}\\
    & the opinion of a blue agent.\\[6pt]
    \hline
    \ms
    \multirow{2}{*}
    {$\left(\rho_1, \rho_2 + \frac{1}{(1-\alpha) N}\right)$}
    & A blue agent in clique $2$ adopts
    & $(1 - \alpha) N (1-\rho_2) \, \cdot$\\
    & the opinion of a red agent.
    & \quad
      $\frac{(1-\alpha) N [(1-\alpha) N - 1] \rho_2 \, + \,
      X \rho_1} 
      {(1 - \alpha) N [(1 - \alpha) N - 1] \, + \, X}$\\[6pt]
    \hline
    \ms
    \multirow{2}{*}
    {$\left(\rho_1, \rho_2 - \frac{1}{(1-\alpha)N}\right)$}
    & A red agent in clique 2 adopts
    & $(1 - \alpha) N \rho_2 \, \cdot$\\
    & the opinion of a blue agent.
    & \quad
      $\frac{(1-\alpha) N [(1-\alpha) N - 1] (1 - \rho_2) \, + \,
      X (1 - \rho_1)}
      {(1-\alpha) N [(1 - \alpha) N - 1] \, + \, X}$\\[6pt]
    \hline
    $(\rho_1, \rho_2)$
    & \multicolumn{2}{c}{Negative sum of all rates above.}\\
    \br
  \end{tabular*}
\end{table}

Applying these simplifications, we can derive the transition rate
matrix $\mathbf{Q}$.
For example, if a blue agent in clique $1$ becomes red, the state
changes from $(\rho_1, \rho_2)$ to $\left(\rho_1 + \frac{1}{\alpha N},
\rho_2\right)$.
This transition occurs with a rate that is the product of the
following two factors.
The first factor is the number of blue agents in clique $1$, which is
equal to $\alpha N (1 - \rho_1)$.
The second factor is the fraction of adjacent agents whose opinion is
red.
Because an agent in clique $1$ is connected to $\alpha N - 1$ agents
in clique $1$ and, on average, to $X/(\alpha N)$ agents in clique $2$,
we obtain the transition rate
\begin{equation*}
  Q\left[(\rho_1, \rho_2),
    \left(\rho_1 + \frac{1}{\alpha N}, \rho_2\right)\right] = \alpha N
  (1-\rho_1) \cdot \frac{(\alpha N - 1) \rho_1 + \frac{X}{\alpha N} \,
    \rho_2}{\alpha N - 1 + \frac{X}{\alpha N}}\ .
\end{equation*}
With similar arguments, we can also deduce the remaining elements of
$\mathbf{Q}$.
In~\tref{tab:Q}, we list all nonzero transition rates.
As is convention, we set the diagonal terms of $\mathbf{Q}$ equal
to the negative sum of all other terms in that
row~\cite{durrett_essentials_2016}: $Q(\boldsymbol{\rho},
\boldsymbol{\rho}) = -\sum_{\boldsymbol{\rho}' \neq \boldsymbol{\rho}}
Q(\boldsymbol{\rho}, \boldsymbol{\rho}')$, where $\boldsymbol{\rho} =
(\rho_1, \rho_2)$.
For our simulations, we apply the exact agent-based update rules of
the voter model and take the exact network topology into account where
the degrees are not the same for all vertices in a clique.
For the analytical solution in~\sref{sec:equation-based}, however, we
resort to the approximations that are implicit in $\mathbf{Q}$.

\section{Simulation results}
\label{sec:simulation}

\begin{figure}
  \begin{center}
    \includegraphics[width=\textwidth]{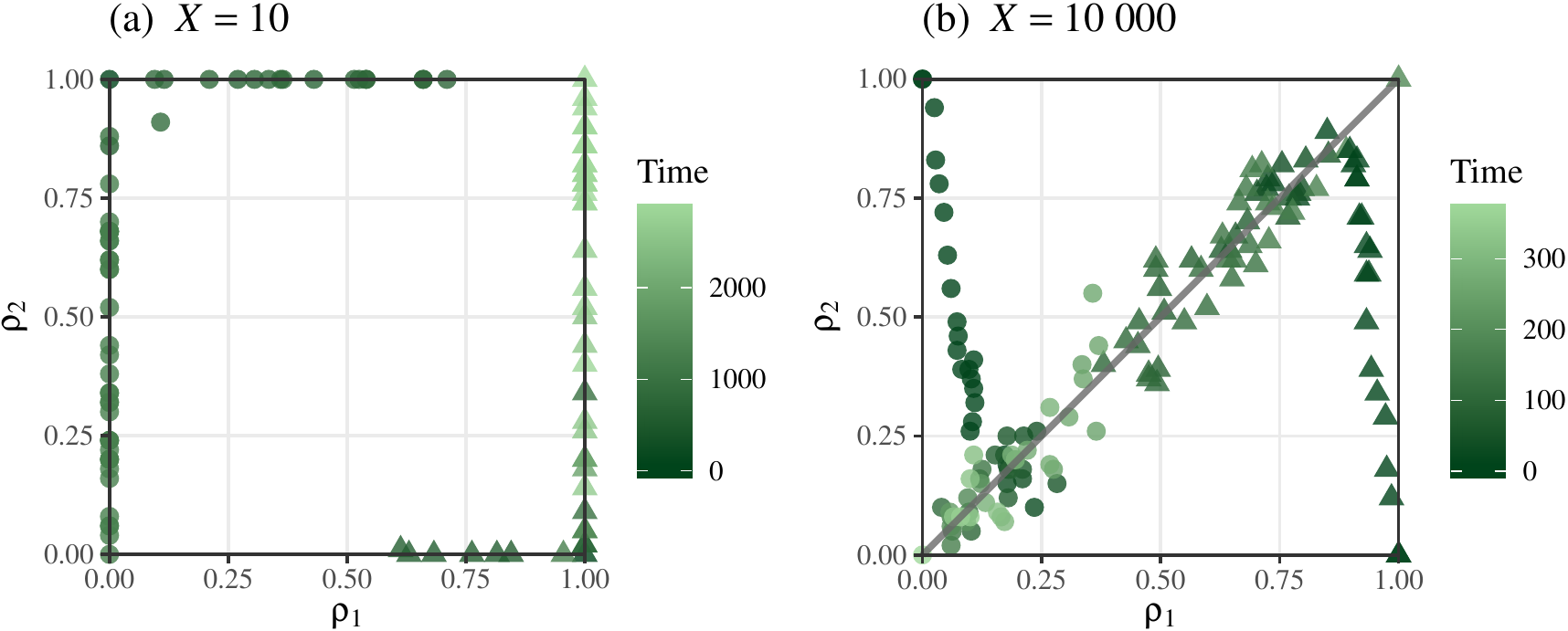}
    \caption{Example trajectories for a two-clique graph with $N =
      500$ vertices and $\alpha = 0.8$ (i.e.\ $400$ vertices are in
      the first clique, $100$ in the second).
      Triangles and circles represent different sample runs of the
      dynamics.
      To avoid overplotting, we only show approximately 50 points
      during each run.
      (a) With only $X = 10$ interclique edges, the states remain
      close to the boundaries of the plot (i.e.\ at least one of the
      cliques is internally almost unanimous).
      (b) If $X = 10\,000$, the system rapidly moves from a state of
      complete polarisation (top-left or bottom-right corner) towards
      the diagonal line ($\rho_1 = \rho_2$).
      The trajectory then remains near the diagonal line until it
      reaches one of the two consensus states in the top-right or
      bottom-left corner of the plot.}
    \label{fig:trajectory}
  \end{center}
\end{figure}

To build intuition about the model, we show 
how the dynamics unfold during several sample runs with $N = 500$ and
$\alpha = 0.8$ in~\fref{fig:trajectory}.
We start the cliques in a state of complete polarisation: within each
clique, opinions are initially unanimous, but there is disagreement
between cliques so that either $(\rho_1, \rho_2) = (1, 0)$ or
$(\rho_1, \rho_2) = (0, 1)$.

In~\fref{fig:trajectory}(a), there are only $X = 10$ interclique
edges; thus, it is difficult for an opinion to invade the clique that
started from the opposite opinion.
Fluctuations occur in only one clique at a time.
Meanwhile, the other clique remains almost unwavering in its support
of its starting opinion.
As a consequence, the trajectory shown in~\fref{fig:trajectory}(a)
mostly remains around the edges of the two-dimensional state space
$(\rho_1, \rho_2) \in [0, 1]^2$.
After a protracted tug of war, confidence in the starting opinion
ultimately vanishes in one of the cliques---usually the smaller one,
with a probability that we will quantify
in equation~\eref{eq:exit_distr_2_clique_voter} below---so that the
system reaches one of the two absorbing states $(0, 0)$ or $(1, 1)$.

By contrast, if $X = 10\,000$, the proportions of red agents $\rho_1$
and $\rho_2$ rapidly approach equality, as shown
in~\fref{fig:trajectory}(b).
Afterwards, the dynamics in one clique almost instantaneously follow
the trends in the other clique so that the trajectory is confined to
the vicinity of the diagonal line $\rho_1 = \rho_2$.
In this case, the cliques behave as one integrated entity despite
being only loosely connected by the network topology.

The distinct behaviours of the model for small and large $X$ lead to
substantially different consensus times, which are evident when
comparing the limits of the colour bar legends in
figures~\ref{fig:trajectory}(a) and~\ref{fig:trajectory}(b).
For $X = \Or(1)$, it can take an extremely long time to reach a
consensus because the cliques hardly exchange any opinions.
If $X \gg 1$, the cliques communicate more frequently with each other
and therefore typically agree on a final opinion sooner.
However, the mean consensus time does not necessarily monotonically
decrease with $X$, as we will now see.

\begin{figure}
  \begin{center}
    \includegraphics[width=\textwidth]{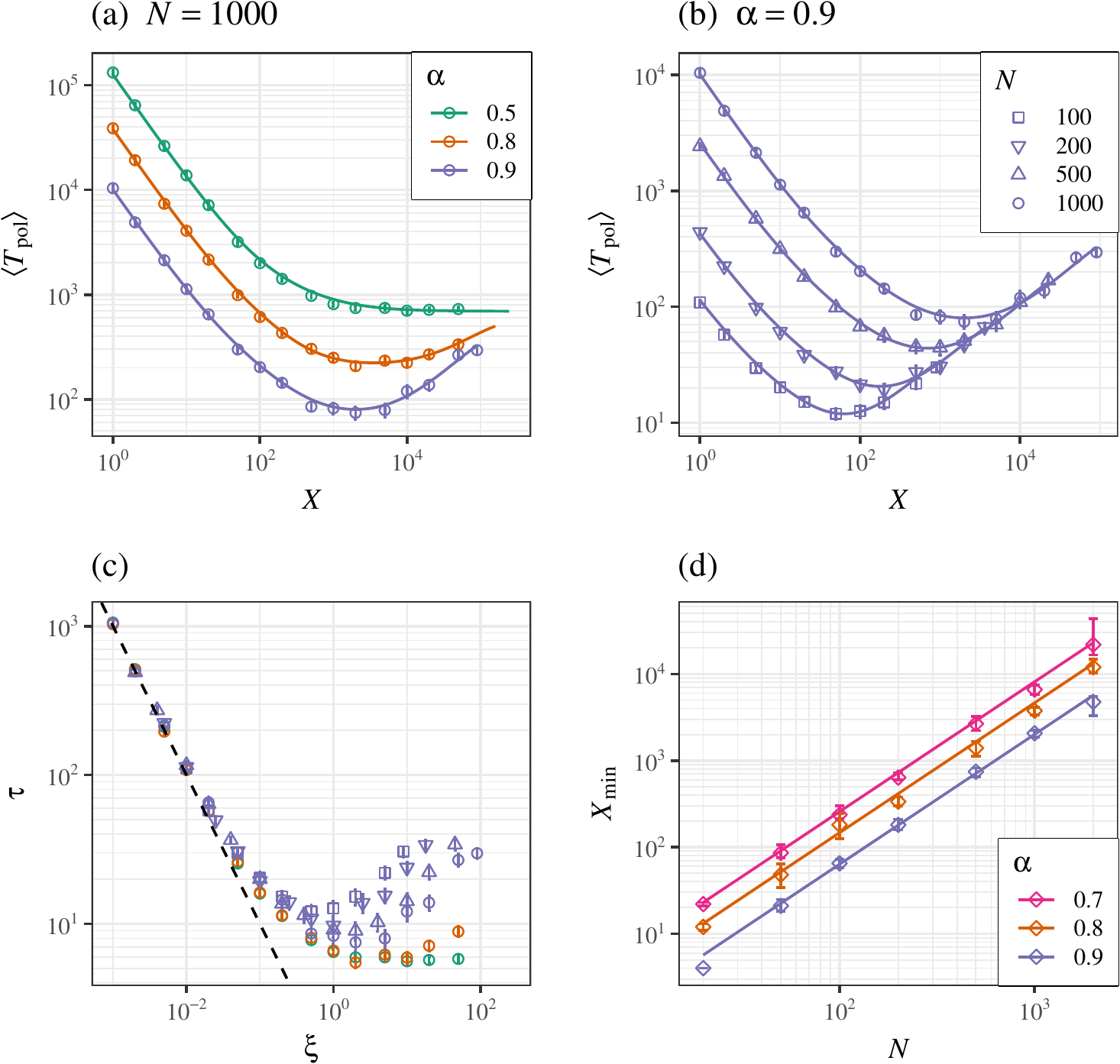}
    \caption{Dependence of the mean consensus time
      $\langle T_{\rm pol} \rangle$ on the number of interclique edges
      $X$.
      Each point represents numerical data from 2000 Monte Carlo simulations
      with a polarised initial condition (i.e.\ cliques are internally
      unanimous, but the cliques disagree with each other).
      Error bars represent 95\% confidence intervals.
      All curves illustrate predictions from the equation-based method
      described in~\sref{sec:equation-based}.
      We emphasise that there are no free parameters in the equations;
      hence, none of the curves shown here needed to be fitted to
      data.
      (a) We fix $N=1000$ and vary $\alpha$.
      (b) We hold $\alpha=0.9$ constant, but $N$ varies.
      (c) The same data as in (a) and (b) but with rescaled
      coordinate axes.
      We plot $\xi = X/N$ along the horizontal axis and
      $\tau = \left(\frac{1}{\alpha^2} +
        \frac{1}{(1 - \alpha)^2}\right)
      \langle T_{\rm pol} \rangle / N$
      along the vertical axis.
      The dashed line represents the reciprocal relationship $\tau =
      1/\xi$.
      (d) The number of edges $X_{\rm min}$ that minimises
      $\langle T_{\rm pol} \rangle$ follows the power law
      $X_{\rm min} \propto N^{3/2}$ predicted
      by equation~\eref{eq:Xmin_2_clique_voter}.
      }
    \label{fig:X}
  \end{center}
\end{figure}

We denote the consensus time from the completely polarised initial
state $(\rho_1, \rho_2) = (1, 0)$ by $T_{\rm pol}$ and its mean by
$\langle T_{\rm pol} \rangle$, which is an average over different
realisations of the stochastic voter-model dynamics and over different
randomly sampled networks with $X$ interclique edges.
In~\fref{fig:X}(a), we show simulation results for $\langle T_{\rm pol}
\rangle$ as a function of $X$ for $N = 1000$ and different values of
$\alpha$.
Because the dynamics for $\alpha$ and $1-\alpha$ are identical if we
exchange the labels of the cliques and opinions, we only plot results
for $\alpha \geq \frac{1}{2}$.
For all values of $\alpha$, $\langle T_{\rm pol} \rangle$ attains its
maximum at $X=1$ and initially decreases as we insert more interclique
edges, consistent with the intuition that more connections lead to a
faster consensus.
Surprisingly, however, if $\alpha \neq \frac{1}{2}$, the trend
reverses as we keep increasing $X$: $\langle T_{\rm pol} \rangle$ passes
through a minimum and then increases again as the network becomes a
complete graph, where $X = \alpha (1 - \alpha) N^2$.
This increase is more pronounced when the difference in
clique sizes is larger.
For $\alpha = 0.9$, we can reduce the mean consensus time by $\approx
76\%$ if we cut $\approx 98\%$ of the interclique ties in the complete
graph.

In~\fref{fig:X}(b), we fix $\alpha = 0.9$ and vary $N$.
In general, an increase in $N$ shifts the curves towards larger values
of $\langle T_{\rm pol} \rangle$ and $X$.
However, the curves' overall shapes remain similar.
The common pattern behind the data plotted in figures~\ref{fig:X}(a)
and~\ref{fig:X}(b) becomes clearer in~\fref{fig:X}(c), where we plot
the rescaled variable
$\tau = \left(\frac{1}{\alpha^2} + \frac{1}{(1 - \alpha)^2}\right)
\langle T_{\rm pol} \rangle / N$ versus $\xi = X / N$.
For $\xi \ll 1$, the rescaled functions collapse onto the same
function $\tau = 1/\xi$.
The scaling relation $\tau \propto 1/\xi$ was pointed out for the
special case $\alpha = \frac{1}{2}$ in~\cite{masuda_voter_2014}.
Our simulations and the equation-based analysis
in~\sref{sec:equation-based} show that $\tau$ and $1/\xi$ are not
merely proportional but equal in the limit $\xi \to 0$.
This result is valid for all clique sizes assuming a completely
polarised initial state.
For other initial conditions, we also find that $\tau \propto 1/\xi$
but with different proportionality factors, which can be calculated
with the method presented in~\sref{sec:equation-based}.

\Fref{fig:X}(d) reveals another emergent scaling relation.
In this scatterplot, the abscissa is the network size $N$.
The ordinate is the number of interclique edges $X_{\rm min}$ that
minimises $\langle T_{\rm pol} \rangle$.
For each combination of $N$ and $\alpha$ in~\fref{fig:X}(d), we
perform 2000 Monte Carlo simulations.
We then estimate $X_{\rm min}$ from the locally estimated scatterplot
smoothing (LOESS) regression curves and establish error bars with
bootstrapping.
For a fixed value of $\alpha$, the data follow the power law
$X_{\rm min} \propto N^{3/2}$.
Thus, to minimise the mean consensus time, the agents must strike a
balance between a sparse and a dense interclique connectivity.
On one hand, the optimal number of interclique links per agent grows
as $k_{{\rm inter}, v} \propto \sqrt{N}$.
On the other hand, the optimal number of interclique edges $X_{\rm
  min}$ is only a vanishing fraction of
the number $\alpha (1 - \alpha) N^2$ of all possible interclique
edges in the limit $N\to\infty$.

In summary, the Monte Carlo simulations reveal three main features
of the two-clique voter dynamics starting from cliques with opposite
opinions.
First, $\langle T_{\rm pol} \rangle$ is a U-shaped function of $X$ as
long as $\alpha \neq \frac{1}{2}$.
Notably, the global minimum does not coincide with a complete
graph.
Second, the mean consensus time obeys the identity $\tau = 1/\xi$ or,
equivalently,
\begin{equation}
  \langle T_{\rm pol} \rangle = \frac{\alpha^2 (1-\alpha)^2}
  {2\alpha^2 - 2\alpha + 1} \frac{N^2}{X}
  \label{eq:T_scaling}
\end{equation}
as long as $X \ll N$.
Third, the number of interclique edges that minimises $\langle
T_{\rm pol}\rangle$ satisfies the scaling relation
$X_{\rm min} \propto N^{3/2}$.
We now demonstrate how these results can be derived from the
transition rates in table~\ref{tab:Q}.

\section{Equation-based analysis}
\label{sec:equation-based}

Let us denote the mean consensus time conditioned on the initial
opinions $(\rho_1, \rho_2)$ by $\langle T(\rho_1, \rho_2) \rangle$.
With this notation,
$\langle T_{\rm pol} \rangle = \langle T(1, 0) \rangle$.
Because $\langle T(\rho_1, \rho_2) \rangle$ is the mean exit time
of a Markov chain with absorbing states $(0, 0)$ and $(1, 1)$, it must
satisfy $\langle T(0, 0) \rangle = \langle T(1, 1) \rangle = 0$
and~\cite{durrett_essentials_2016}
\begin{equation*}
  \sum_{y, z} Q[(\rho_1, \rho_2), (y, z)] \langle T(y, z)
  \rangle = -1
  \label{eq:general_exit_time}
\end{equation*}
if $(\rho_1, \rho_2) \not\in \{(0, 0), (1, 1)\}$.
Next, we insert $\mathbf{Q}$ from table~\ref{tab:Q} and perform, in
the parlance of mathematical population genetics, a diffusion
approximation~\cite{ewens_mathematical_2004}: we assume $N \gg 1$ and
take the continuum limit.
The result is the partial differential equation
\begin{equation}
  \fl
  \eqalign{
    \frac{
    (\alpha N - 1) \rho_1 (1 - \rho_1) +
    \frac{X}{2 \alpha N} (\rho_1 + \rho_2 - 2 \rho_1 \rho_2)
    }{\alpha N (\alpha N - 1) + X} 
    \frac{\partial^2 \langle T \rangle}{\partial \rho_1^2}\\
    \quad
    + \frac{
    [(1 - \alpha) N - 1] \rho_2 (1 - \rho_2)
    + \frac{X}{2 (1 - \alpha) N}
    (\rho_1 + \rho_2 - 2 \rho_1 \rho_2)
    }{(1 - \alpha) N [(1 - \alpha) N - 1] + X} 
    \frac{\partial^2 \langle T \rangle}{\partial \rho_2^2}\\
    \quad
    + \frac{X}{\alpha N (\alpha N - 1) + X} (\rho_2 - \rho_1)
    \frac{\partial \langle T \rangle}{\partial \rho_1}\\
    \quad
    + \frac{X}{(1 - \alpha) N [(1 - \alpha) N -1] + X}
    (\rho_1 - \rho_2)
    \frac{\partial \langle T \rangle}{\partial \rho_2} + O (N^{-2})\\
    = -1.
  }
  \label{eq:general_exit_time_pde_2_clique}
\end{equation}
Finding an exact solution to
equation~\eref{eq:general_exit_time_pde_2_clique} would be a
formidable task, but the result would not be directly useful.
Instead, we aim for an approximate solution.
First, we find a solution that is valid if $X = \Or(N)$.
Afterwards, we derive an approximation for the case where $X \gg N$.
Finally, we interpolate between these two approximations to arrive at
a solution that fits the data remarkably well over the entire range
from $X = 1$ to the complete graph with $X = \alpha (1 - \alpha) N^2$.
The solid curves in~\fref{fig:X} are based on this interpolation.

\subsection{Approximate solution if $X = \Or(N)$}

For a sparse interclique connectivity, the leading terms of
equation~\eref{eq:general_exit_time_pde_2_clique} up to and including
$\Or\left(N^{-1}\right)$ are
\begin{equation}
  \eqalign{
    & \frac{1}{\alpha N}\, \rho_1 (1-\rho_1)
    \frac{\partial^2 \langle T\rangle} {\partial \rho_1^2} +
    \frac{1}{(1-\alpha) N} \rho_2 (1-\rho_2)
    \frac{\partial^2 \langle T\rangle} {\partial \rho_2^2}\\
    & \quad + \frac{X} {\alpha^2 N^2}\, (\rho_2 - \rho_1)
    \frac{\partial \langle T\rangle} {\partial \rho_1} +
    \frac{X}{(1-\alpha)^2 N^2} (\rho_1 - \rho_2)
    \frac{\partial \langle T\rangle} {\partial \rho_2} = -1.
  }
  \label{eq:pde_T_2_clique}
\end{equation}
We are not aware of an exact solution to
equation~\eref{eq:pde_T_2_clique}, but  we assume that it can be
expressed as a power series.
The main features already become apparent when only expanding up to
the quadratic terms.
We denote this approximation by $t_{\rm sparse}$ to indicate that
this expression is valid if we only have a sparse connectivity between
cliques:
\begin{equation}
  t_{\rm sparse}(\rho_1, \rho_2) =
  \sum_{i=0}^2 \sum_{j=0}^2 c_{i, j} \left(\rho_1 - \frac{1}{2}\right)^i
  \left(\rho_2 - \frac{1}{2}\right)^j\ .
  \label{eq:T_sparse_2_clique_voter}
\end{equation}
Because $\langle T \rangle$ is symmetric with respect to
$\left(\frac{1}{2}, \frac{1}{2}\right)$, we must have $c_{i, j} = 0$
if either $i$ is odd and $j$ is even or vice versa.
Only five coefficients remain that can possibly be nonzero: $c_{0,
  0}$, $c_{0, 2}$, $c_{1, 1}$, $c_{2, 0}$, and $c_{2, 2}$.
We can determine these coefficients from
equation~\eref{eq:pde_T_2_clique} and the boundary conditions.
We skip the details here and instead refer to the appendix, where we
show that
\begin{equation}
  \fl
  \eqalign{
    \langle T_{\rm pol} \rangle
    & \approx
    t_{\rm sparse}(1, 0) =
    -\frac{1}{2} c_{1, 1}\\
    & = \frac{\alpha^2 (1-\alpha)^2 N}{X d(\alpha, N, X)}
    \big[
    2(2\alpha^2 - 2\alpha + 1) X^3 + 2(\alpha^2 - \alpha + 1)
    N X^2\\ 
    & \hspace{3cm} + \alpha (1-\alpha) (2\alpha^2 - 2\alpha + 3) N^2 X +
    \alpha^2 (1-\alpha)^2 N^3
    \big]
  }
  \label{eq:Tpol_sparse_2_clique_voter}
\end{equation}
with the auxiliary function
\begin{equation}
  \eqalign{
    d(\alpha, N, X) =
    & (3\alpha^2 - 3\alpha + 1)(2\alpha^2 - 2\alpha + 1) X^2\\
    & + \alpha(1-\alpha) (4\alpha^4 - 8\alpha^3 + 11\alpha^2
    - 7\alpha + 2) N X\\
    & + \alpha^2 (1-\alpha)^2 (2\alpha^2-2\alpha + 1) N^2
  }
  \label{eq:d}
\end{equation}
in the denominator.

\begin{figure}
  \begin{center}
    \includegraphics[width=0.55\textwidth]{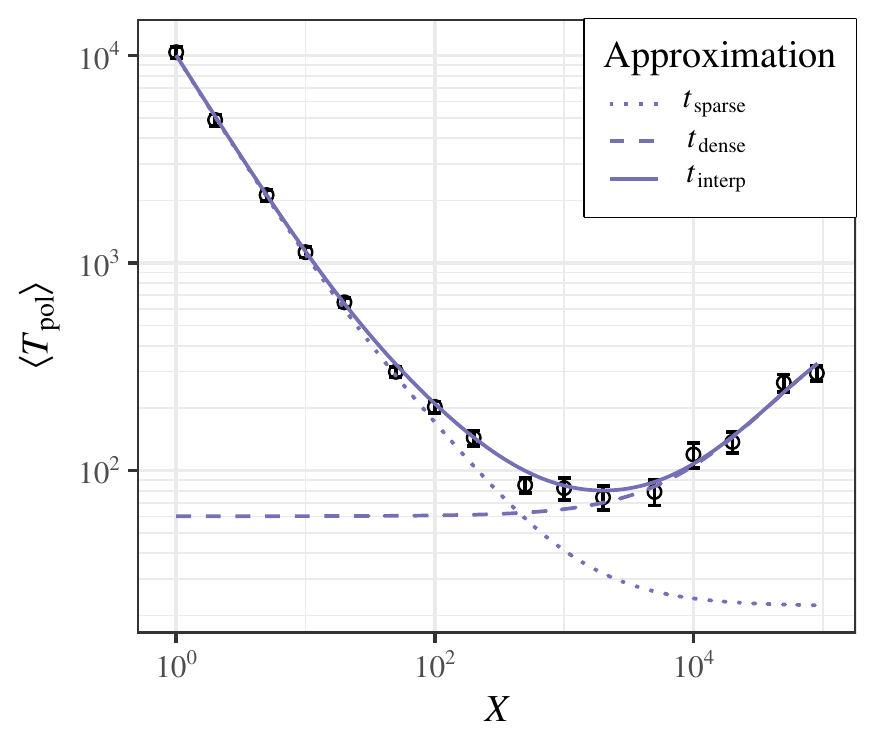}
  \end{center}
  \caption{Illustration of the approximations for $\langle T_{\rm pol}
    \rangle$ presented in~\sref{sec:equation-based}.
    As an example, we show data for $N=1000$ and $\alpha=0.9$ as open
    circles.
    Each circle represents the mean of 2000 Monte Carlo simulations. 
    The error bars are 95\% confidence intervals.
    The approximation $t_{\rm sparse}$, given
    by~\eref{eq:Tpol_sparse_2_clique_voter} and shown as a dotted
    line, fits the data well if the cliques are sparsely connected
    (i.e.\ $X < N$) but loses accuracy if we insert more edges
    between cliques.
    Equation~\eref{eq:Tdense_2_clique_voter} presents the alternative
    approximation $t_{\rm dense}$ (dashed line), which is a much
    better estimate than $t_{\rm sparse}$ if $X > N$ but is worse for
    a sparse interclique connectivity.
    In~\eref{eq:Tinterp_2_clique_voter}, we define an interpolation
    $t_{\rm interp}$ that asymptotically behaves like $t_{\rm sparse}$
    for small $X$ and $t_{\rm dense}$ for large $X$.
    Even for intermediate $X$, $t_{\rm interp}$ closely approximates
    the simulation data (solid line).
  }
  \label{fig:interpolation}
\end{figure}

In~\fref{fig:interpolation}, we compare the numerical data for
$\alpha=0.9$ and $N=1000$ with the approximation in
equation~\eref{eq:Tpol_sparse_2_clique_voter}, shown as a dotted
line.
In the limit $X/N \to 0$, $t_{\rm sparse}$ is an excellent fit because
the asymptotic behaviour of
equation~\eref{eq:Tpol_sparse_2_clique_voter} is consistent with
equation~\eref{eq:T_scaling}.
For large $X$, however, equation~\eref{eq:Tpol_sparse_2_clique_voter}
predicts a consensus time that is too short.
To resolve this problem, we now derive an approximation that is more
suitable if $X$ is large.

\subsection{Approximate solution if $X \gg N$}
\label{subsec:XggN}

The probability of reaching a red consensus from the initial condition
$(\rho_1, \rho_2)$ is, in general, given by the martingale
$m(\rho_1, \rho_2)$ that satisfies $m(0, 0) =  0$, $m(1, 1) = 1$
and~\cite{durrett_essentials_2016}
\begin{equation*}
  \sum_{y, z} Q[(\rho_1, \rho_2), (y, z)] m(y, z) = 0
\end{equation*}
for all $(\rho_1, \rho_2) \not\in \{(0, 0), (1, 1)\}$.
By inserting the formulae for the elements of $\mathbf{Q}$ from
table~\ref{tab:Q}, we can verify that the solution is
\begin{equation}
  \fl
  m(\rho_1, \rho_2) =
  \frac{(\alpha^2 N^2 - \alpha N + X) \rho_1 +
    [(1-\alpha)^2 N^2 - (1-\alpha) N + X] \rho_2}
  {(2\alpha^2 - 2\alpha + 1) N^2 - N + 2X}\ .
  \label{eq:exit_distr_2_clique_voter}
\end{equation}
This result is valid regardless of whether $X$ is small or large. 

If the cliques are densely connected, we have seen in
figure~\ref{fig:trajectory}(b) that we can assume that $\rho_1 =
\rho_2$ after a short transient.
Similar adiabatic approximations have been applied, e.g.\
in~\cite{sood_voter_2005, sood_voter_2008-1,
  constable_population_2014, constable_fast-mode_2014,
  gastner_consensus_2018-1}.
By inserting $\rho_1 = \rho_2$
into equation~\eref{eq:exit_distr_2_clique_voter}, it follows that the
fraction of red agents in each clique is equal to $m$.
Thus, we can substitute $m$ for $\rho_1$ and $\rho_2$
in equation~\eref{eq:pde_T_2_clique}.
Bearing in mind that
\begin{equation*}
\frac{\partial^2 \langle T \rangle}{\partial \rho_i^2}
= \frac{\partial^2 m}{\partial \rho_i^2}
\frac{d \langle T \rangle}{dm} +
\left(\frac{\partial m}{\partial \rho_i}\right)^2
\frac{d^2 \langle T \rangle}{dm^2}
\end{equation*}
for $i \in \{1, 2\}$ and keeping only the leading-order terms, we
obtain the following second-order ordinary differential equation:
\begin{equation}
  \frac{
    \alpha (1 - \alpha) N^2 [(3\alpha^2 - 3\alpha + 1) N^2 + 2X] + X^2
  }{
    \alpha (1 - \alpha) N [(2\alpha^2 - 2\alpha + 1) N^2 + 2X]^2
  }
  m (1-m) \frac{d^2 \langle T \rangle}{dm^2} = -1.
  \label{eq:Tm_2_clique_voter}
\end{equation}

We call the solution to equation~\eref{eq:Tm_2_clique_voter} $t_{\rm
  dense}$, where the subindex `dense' expresses that the equation is
derived under the assumption that $X \gg N$.
The absorbing boundary condition
\begin{equation*}
  t_{\rm dense}(m = 0) = t_{\rm dense}(m = 1) = 0
\end{equation*}
uniquely determines the solution
\begin{equation}
  \fl
  \eqalign{
    t_{\rm dense}(m) =\\
    \quad
    -\frac{
      \alpha (1 - \alpha) N [(2\alpha^2 - 2\alpha + 1) N^2 + 2X]^2
    }{
      \alpha (1 - \alpha) N^2 [(3\alpha^2 - 3\alpha + 1) N^2 + 2X]
      + X^2
    }[m \ln m + (1 - m) \ln(1 - m)].
  }
  \label{eq:Tdense_2_clique_voter}
\end{equation}
\Fref{fig:interpolation} confirms that $t_{\rm dense}$ fits the data
from the Monte Carlo simulations in the range $X \gg N$.
In particular, $t_{\rm dense}$ correctly predicts an increasing
mean consensus time for large $X$.
A closer look at equation~\eref{eq:Tdense_2_clique_voter} reveals that
$t_{\rm dense}$ increases because the minority opinion gains a
slightly higher probability of winning.
For a polarised initial condition on a network with $N = 1000$ and
$\alpha = 0.9$, we find that $m(1, 0) \approx 0.98$ if $X=1$, but
$m(1, 0) = 0.9$ if the graph is complete (i.e.\ $X = 90\,000$).
Hence, the blue minority increases its probability of winning from
2\% to 10\%.
At first glance, the difference in $m$ may seem to be small, but its
effect is amplified by the nearby singularity of the function
$\ln(1 - m)$, which appears on the right-hand side of
equation~\eref{eq:Tdense_2_clique_voter}.
As a consequence, $t_{\rm dense}$ increases by a factor of
approximately $5.4$ as $X$ increases from $1$ to $90\,000$.
We conclude that networks designed for a fast consensus must strike a
compromise between two opposing trends.
On one hand, frequent opinion exchanges between the cliques are
necessary to quickly agree on the same opinion.
On the other hand, additional interclique edges give the minority
clique greater influence, causing more self-doubt within the
majority clique and consequently slower convergence towards a shared
opinion.

\subsection{Derivation of $X_{\rm min} \propto N^{3/2}$}

While $t_{\rm dense}$ is an excellent approximation of the simulated
data if $X \gg N$, it unfortunately underestimates the true value of
$\langle T \rangle$ in the range $X < N$.
In this sense, $t_{\rm dense}$ is the opposite of $t_{\rm sparse}$
from~\sref{subsec:XggN}: we found that $t_{\rm sparse}$ approximates
$\langle T \rangle$ well for small $X$ but substantially deviates 
for large $X$ (\fref{fig:interpolation}).
To obtain the benefits of both $t_{\rm dense}$ and $t_{\rm sparse}$
but none of their disadvantages, we construct an interpolation
$t_{\rm interp}$ as follows. 
We first add $t_{\rm dense}$ and $t_{\rm sparse}$
and then subtract the asymptotic value of $t_{\rm sparse}$
in the limit of a dense interclique connectivity:
\begin{equation}
  \fl
  \eqalign{
    t_{\rm interp}(\rho_1, \rho_2, X) =\\
    \quad
    t_{\rm dense}[m(\rho_1, \rho_2, X), X]
    + t_{\rm sparse}(\rho_1, \rho_2, X)
    - \lim_{X' \to \infty} t_{\rm sparse}(\rho_1, \rho_2, X'),
  }
  \label{eq:Tinterp_2_clique_voter}
\end{equation}
where we have explicitly included $X$ among the independent
variables.
This interpolation approximates the true value of $\langle T \rangle$
in the limit of a minimal or maximal interclique connectivity and is
also an excellent approximation for all intermediate values of $X$.
The solid curves in figures~\ref{fig:X}(a), \ref{fig:X}(b),
and~\ref{fig:interpolation} confirm that $t_{\rm interp}$ fits well
for all $X$.

Equipped with an approximation of $\langle T \rangle$, we can now
determine how many edges must be inserted between polarised cliques to
minimise the mean consensus time.
From equation~\eref{eq:Tinterp_2_clique_voter} and the condition
$\partial t_{\rm interp} / \partial X = 0$ for the minimum, it follows
that we are looking for the solution $X_{\rm min}$ of the equation
\begin{equation}
  \frac{\partial t_{\rm sparse}(1, 0, X)}{\partial X} =
  -\frac{\partial t_{\rm dense}(1, 0, X)}{\partial X}\ .
  \label{eq:Xmin_condition_2_clique_voter}
\end{equation}
To simplify the calculation, let us assume that $X_{\rm min}$
increases between linearly and quadratically in $N$.
Expressed in formal notation, we assume that $1 \ll N = o(X_{\rm
  min})$ and $X_{\rm min} = o\left(N^2\right)$.
In this case, we can expand $t_{\rm sparse} / N$ and $t_{\rm dense} /
N$ as Taylor series in terms of $N / X$ and $X / N^2$, respectively.
Rearranging equations~\eref{eq:Tpol_sparse_2_clique_voter}
and~\eref{eq:Tdense_2_clique_voter}, we find that
\begin{eqnarray}
  \fl
  \eqalign{
  \frac{t_{\rm sparse}(1, 0, X)}{N} =
  & \frac{2 \alpha^2 (1-\alpha)^2}{3\alpha^2 - 3\alpha + 1}\\
  & + \frac{2 \alpha^2 (1-\alpha)^2 (2\alpha^4
    - 4\alpha^3 + 6\alpha^2 - 4\alpha +1)}
    {(3\alpha^2 - 3\alpha + 1)^2}
    \frac{N}{X} + O\left[\left(\frac{N}{X}\right)^2\right],
    }\\
  \fl
  \eqalign{
  \frac{t_{\rm dense}(1, 0, X)}{N} =
  & \frac{2 \alpha^2 - 2 \alpha + 1}{3 \alpha^2 - 3 \alpha + 1}
    [(2 \alpha^2 - 2 \alpha + 1) \ln(2 \alpha^2 - 2 \alpha + 1)\\
  & \hspace{2.5cm} - 2 \alpha^2 \ln(\alpha) - 2(1 - \alpha)^2 \ln(1
    - \alpha)]\\
  & + \frac{2 (1 - 2\alpha)}{(3 \alpha^2 - 3 \alpha + 1)^2}
    [(4 \alpha^3 - 5 \alpha^2 + 3 \alpha - 1)
    \ln(\alpha)\\
  & \hspace{3.5cm} + (4 \alpha^3 - 7 \alpha^2 + 5\alpha -
    1)
    \ln(1 - \alpha)\\
  & \hspace{3.5cm} + (1 - 2 \alpha) (2 \alpha^2 - 2 \alpha + 1)
    \ln(2 \alpha^2 - 2 \alpha + 1)] \frac{X}{N^2}\\
  & + O\left[\left(\frac{X}{N^2}\right)^2\right].
    }
    \label{eq:Tdense_taylor_2_clique_voter}
\end{eqnarray}
We now combine
equations~\eref{eq:Xmin_condition_2_clique_voter}--\eref{eq:Tdense_taylor_2_clique_voter}
and drop the higher-order terms.
The result is
\begin{equation}
  X_{\rm min} = \alpha (1-\alpha) \sqrt{\frac{2\alpha^4
      - 4\alpha^3 + 6\alpha^2 - 4\alpha +1}{q(\alpha)}}\,
  N^{3/2}
  \label{eq:Xmin_2_clique_voter}
\end{equation}
with
\begin{equation*}
  \eqalign{
  q(\alpha) = (1-2\alpha)\,[
  & (4 \alpha^3 - 5 \alpha^2 + 3 \alpha - 1)
    \ln(\alpha)\\
  & + (4 \alpha^3 - 7 \alpha^2 + 5\alpha -
    1) \ln(1 - \alpha)\\
  & + (1 - 2 \alpha) (2 \alpha^2 - 2 \alpha + 1)
  \ln(2 \alpha^2 - 2 \alpha + 1)].
  }
\end{equation*}
\Fref{fig:X}(d) confirms that the predicted $X_{\rm min}$ (straight
line) is in good agreement with the simulation data.

\section{Discussion}
\label{sec:discussion}

In this article, we have studied the voter model for one of the
simplest types of community structure: two cliques connected by a
fixed number of edges. Previously, equations were only
available for the special case of two equally large cliques.
Even for this special case, only the asymptotic behaviour for either
an extremely sparse or extremely dense interclique connectivity was
known~\cite{masuda_voter_2014}.
Here, we have introduced a heterogeneous mean-field approximation and
an interpolation technique that allow us to treat cliques of unequal
sizes.
Furthermore, equation~\eref{eq:Tinterp_2_clique_voter} makes a
prediction for the mean consensus time that goes beyond a mere scaling
law with an unknown proportionality constant.
Instead, we can calculate concrete numbers that are in excellent
agreement with Monte Carlo simulations for any number of
interclique edges, $X$, including cases where the adiabatic
approximation at the heart of~\cite{sood_voter_2005,
  sood_voter_2008-1, constable_population_2014,
  constable_fast-mode_2014, gastner_consensus_2018-1} fails.
In particular, equation~\eref{eq:Xmin_2_clique_voter} predicts the
number of interclique edges, $X_{\rm min}$, necessary to minimise the
mean consensus time.
Our derivation of equation~\eref{eq:Xmin_2_clique_voter} reveals that,
at the optimum, the smaller clique must be exposed to the majority
opinion, but we must not allow the smaller clique to influence the
larger clique too strongly.
The result $X_{\rm min}\propto N^{3/2}$ exemplifies how our
methodology is able to answer a sociological question with a specific
and surprising quantitative prediction.

We have considered the scenario where the cliques have different
sizes, consistent with empirical observations that community size
distributions tend to be highly
heterogeneous~\cite{lancichinetti_characterizing_2010,
  yang_structure_2014}.
Still, real community structures are considerably more complex than
our model.
For example, communities in real networks are typically much sparser
than cliques~\cite{lancichinetti_characterizing_2010}.
Moreover, real communities are not necessarily as clearly separated
from each other as in our model.
Instead, the boundaries between communities are often fuzzy so that
vertices can often not be uniquely attributed to a single
community~\cite{yang_structure_2014}.
Even if communities do not overlap, it is highly restrictive to
assume that their number is exactly equal to two.

Besides assuming a stylised network topology, we have also applied a
particularly simple update rule.
In our model, agents can choose between only two different opinions,
which must be truthfully signalled to all neighbours.
A more sophisticated model may distinguish between private and
publicly displayed opinions~\cite{gastner_consensus_2018-1}, thereby
giving agents the opportunity to be hypocrites (i.e.\ they may
represent an opinion in public that is contrary to their inner
belief)~\cite{gastner_impact_2019}.
If there are more than two possible opinions, yet more complex update
rules are conceivable~\cite{vazquez_ultimate_2004-1}.
Further potential model variants include zealots who never change their
opinions~\cite{mobilia_does_2003-1, mobilia_role_2007-1, bhat_opinion_2019}
or agents who query more than a single neighbour before switching
opinions~\cite{lambiotte_dynamics_2007-1, castellano_nonlinear_2009}.
Updates may happen simultaneously instead of
asynchronously~\cite{gastner_ising_2015}.
The distribution of waiting times between updates may be more
right-skewed than an exponential
distribution~\cite{takaguchi_voter_2011,
  fernandez-gracia_timing_2013}.
There may even be different waiting time distributions for different
agents~\cite{masuda_heterogeneous_2010-1} or different rates of
opinion exchange along different
edges~\cite{baronchelli_voter_2011}.
These and many more modifications of the basic voter model have been
previously studied~\cite{redner_reality-inspired_2019}.
It would be interesting to investigate how the two-clique topology
influences the dynamics in these cases.

The voter model is not only relevant in the context of opinions in
social networks.
It can also be interpreted as a model for language
evolution~\cite{castello_ordering_2006, blythe_stochastic_2007}, where
the state of an agent is a linguistic token instead of an opinion.
In this context, a two-clique topology may represent a society that is
split into two groups because of geography (e.g.\ a language island
separated from the mainland).
While a quick consensus may be preferable in the context of opinion
formation, the extinction of language variants is a cultural loss that
should be avoided or at least delayed.
Because the deliberate removal of interclique edges can hardly be
socially desirable, our model suggests that the best way to extend the
lifetime of a language variant is to increase the size of the
minority clique.

Even before the voter model appeared in the sociological and physics
literature, it had been introduced in biology, albeit under different
names.
For example, the Moran process represents the spread of alleles in a
population with a model that is---at least for the panmictic
population considered in Moran's 1958
paper~\cite{moran_random_1958}---equivalent to the voter model.
Other biologists have interpreted the two-dimensional voter model
as a competition for territory between
species~\cite{clifford_model_1973-1, chave_spatial_2001}.
From a biological perspective, the voter model on a network with
two communities may be viewed at first glance as a direct
implementation of Wright's island model, where `the
total population is assumed to be divided into subgroups, each
breeding at random within itself, except for a certain proportion of
migrants'~\cite{wright_isolation_1943}.
Still, there is a subtle but important difference between the voter
model and the Moran process (also known as the invasion
process~\cite{sood_voter_2008-1}).
When interpreted in the context of opinion formation, it makes sense
to assume that the focal agent adopts the opinion of a random
neighbour.
In biology, by contrast, the interaction between the focal vertex and
its neighbour is usually in the opposite direction: the offspring of
the focal vertex spreads the parent's state to a neighbouring site.
On a degree-regular network (e.g.\ a complete graph, as in Moran's
paper~\cite{moran_random_1958}), both update rules lead to the same
stochastic process.
For heavy-tailed degree distributions, however, the two update rules
are known to result in substantially different
dynamics~\cite{castellano_effect_2005, sood_voter_2005,
  sood_voter_2008-1}.
The degree distribution of a two-clique network is not heavy-tailed
but bimodal with peaks at $\alpha N + X / (\alpha N)$ and
$(1-\alpha) N + X / [(1 - \alpha) N]$.
Whether this topology causes a difference between the voter model and
the Moran process is a question for future research.
The methodology we have presented in this article opens the door to
such studies of voter-like models for networks with a community
structure.

\ack
We are grateful to Be\'ata Oborny for helpful comments on this
manuscript.
This work was supported by the Singapore Ministry of Education and
a Yale-NUS College start-up grant (R-607-263-043-121).
We would like to thank Editage (www.editage.com) for English language
editing.

\appendix
\setcounter{section}{1}

\section*{Appendix. Derivation of
  equation~\eref{eq:Tpol_sparse_2_clique_voter}}

We are looking for a solution to equation~\eref{eq:pde_T_2_clique} by
expanding $\langle T \rangle$ as a Taylor series:
\begin{equation}
\langle T(\rho_1, \rho_2) \rangle = \sum_{i=0}^\infty
  \sum_{j=0}^\infty c_{i, j} \left(\rho_1 - \frac{1}{2}\right)^i
  \left(\rho_2 - \frac{1}{2}\right)^j,
  \label{eq:T_series_2_clique}
\end{equation}
where the coefficients $c_{i, j}$ can be determined
from equation~\eref{eq:pde_T_2_clique} and the boundary conditions.
In equation~\eref{eq:T_series_2_clique}, we have chosen to expand
$\langle T \rangle$ around
$(\rho_1, \rho_2) = \left(\frac{1}{2}, \frac{1}{2}\right)$ because it
is a point of symmetry: the dynamics remain identical if we
interchange red and blue opinions.
For this reason, we must have $c_{i, j} = 0$ whenever either $i$ is
odd and $j$ is even or vice versa.
Hence,
\begin{eqnarray*}
  \fl
  \langle T(\rho_1, \rho_2) \rangle =
  c_{0, 0}
  + c_{2, 0} \left(\rho_1 - \frac{1}{2}\right)^2
  + c_{1, 1} \left(\rho_1 - \frac{1}{2}\right)
   \left(\rho_2 - \frac{1}{2}\right)
  + c_{0, 2} \left(\rho_1 - \frac{1}{2}\right)^2\\
  + c_{2, 2} \left(\rho_1  - \frac{1}{2}\right)^2
  \left(\rho_2 - \frac{1}{2}\right)^2
  +\; \textrm{higher-order~terms}.
\end{eqnarray*}
We drop the higher-order terms (i.e.\ those with exponents $\geq 3$)
and denote this approximation of $\langle T\rangle$ by
$t_{\rm sparse}$, just as we did in
equation~\eref{eq:T_sparse_2_clique_voter}.
Inserting $t_{\rm sparse}$ into~\eref{eq:pde_T_2_clique} and comparing
the constant terms on the left- and right-hand sides of the equation,
we find that
\begin{equation}
  (1 - \alpha) \, c_{2, 0} + \alpha \, c_{0, 2}
  = -2 \alpha (1 - \alpha) N.
  \label{eq:T_sparse_inside}
\end{equation}
The blue consensus $(\rho_1, \rho_2) = (0, 0)$ is an absorbing state;
therefore, we demand that $t_{\rm sparse}(0, 0) = 0$ or, equivalently,
\begin{equation}
  16 c_{0, 0}
  + 4 \left(c_{2, 0} + c_{1, 1} + c_{0, 2}\right)
  + c_{2, 2}
  = 0.
  \label{eq:T_sparse_absorbing_corner}
\end{equation}
In the polarised corner of the state space with the coordinates
$(\rho_1, \rho_2) = (1, 0)$ and along the boundaries $(\rho_1, 0)$ and
$(0, \rho_2)$ with $\rho_1 \not\in \{0, 1\}$ and
$\rho_2 \not\in \{0, 1\}$, respectively, we find similar identities
for the coefficients $c_{i, j}$ by
evaluating equation~\eref{eq:pde_T_2_clique}.
Combining these identities with equations~\eref{eq:T_sparse_inside}
and~\eref{eq:T_sparse_absorbing_corner}, we reach the following system
of linear equations:
\begin{eqnarray*}
  \fl
  \left(
  \begin{array}{ccccc}
    0
    & 1 - \alpha
    & 0
    & \alpha
    & 0 \\ 
    16
    & 4
    & 4
    & 4
    & 1\\
    0
    & -4(1 - \alpha)^2 X
    & 2(2 \alpha^2 - 2 \alpha + 1) X
    & -4 \alpha^2 X
    & -(2 \alpha^2 - 2 \alpha + 1) X\\
    0
    & 4 \alpha (1 - \alpha)^2 N
    & 2(1 - \alpha)^2 X
    & -4 \alpha^2 X
    & \alpha (1 - \alpha)^2 N\\
    0
    & -4(1 -  \alpha)^2 X
    & 2 \alpha^2 X
    & 4 \alpha^2 (1 - \alpha) N
    & \alpha^2 (1 - \alpha) N
  \end{array}
      \right)
      \left(
      \begin{array}{c}
        c_{0,0}\\
        c_{0,2}\\
        c_{1,1}\\
        c_{2,0}\\
        c_{2, 2}
      \end{array}
  \right)
  \nonumber\\
  = - 2 \alpha (1 - \alpha) N
  \left(
  \begin{array}{c}
    1\\
    0\\
    2 \alpha (1 - \alpha) N\\
    4 \alpha (1 - \alpha) N\\
    4 \alpha (1 - \alpha) N
  \end{array}
  \right)\ .
\end{eqnarray*}
Its solution is
\begin{eqnarray*}
  \fl
  c_{0,0}
  = \frac{N}{4 X d(\alpha, N, X)}
  \big[
  & 2(2\alpha^2 - 2\alpha + 1)^3 X^3 +
  \alpha (1-\alpha) (\alpha^2-\alpha+1) (9\alpha^2 - 9\alpha + 4)
  N X^2\\
  & - 2\alpha^2 (1-\alpha)^2 (\alpha^2 - \alpha - 1)
  (\alpha^2 - \alpha + 1) N^2 X +
  2 \alpha^4 (1-\alpha)^4 N^3
  \big],
\end{eqnarray*}
\begin{eqnarray*}
  \fl
  c_{2,0}
  = - \frac{\alpha^4 N}{d(\alpha, N, X)}
  \big[
  & 2(2\alpha^2-2\alpha+1) X^2 +
  2(1-\alpha)(2\alpha^2-\alpha+1) N X\\
  & + (1-\alpha)^2 (2\alpha^2 - \alpha + 1) N^2\big],
\end{eqnarray*}
\begin{eqnarray}
  \fl
  c_{1,1} = -\frac{2\alpha^2 (1-\alpha)^2 N}{X d(\alpha, N, X)}
  \big[
  & 2(2\alpha^2 - 2\alpha + 1) X^3 + 2(\alpha^2 - \alpha + 1)
    N X^2
    \label{eq:c11_2_clique_voter}
  \\ 
  & + \alpha (1-\alpha) (2\alpha^2 - 2\alpha + 3) N^2 X +
    \alpha^2 (1-\alpha)^2 N^3
    \big],
    \nonumber
\end{eqnarray}
\begin{eqnarray*}
  \fl
  c_{0,2}
  = -\frac{(1-\alpha)^4 N}{d(\alpha, N, X)}
  \big[
  & 2(2\alpha^2 - 2\alpha + 1) X^2 +
    2\alpha (2\alpha^2 - 3\alpha + 2) N X\\
  & + \alpha^2 (2\alpha^2 - 3\alpha + 2) N^2
    \big],
\end{eqnarray*}
\begin{equation*}
  \fl
  c_{2,2}
  = -\frac{4\alpha^2 (1-\alpha)^2 N^2}{d(\alpha, N, X)}
  \left[
    (3\alpha^2 - 3\alpha + 1) X +
    \alpha (1-\alpha) (2\alpha^2 - 2\alpha + 1) N
  \right],
\end{equation*}
where $d(\alpha, N, X)$ is given by equation~\eref{eq:d}.

In the special case of a polarised initial condition,
equation~\eref{eq:T_series_2_clique} can be simplified thanks to
equation~\eref{eq:T_sparse_absorbing_corner}:
\begin{equation}
  \eqalign{
    \langle T_{\rm pol} \rangle
    & \approx t_{\rm sparse}(1, 0)
    = c_{0, 0} +
    \frac{1}{4} \left(c_{0, 2} - c_{1, 1} + c_{2, 0}\right) +
    \frac{1}{16} c_{2, 2}\\
    & = -\frac{1}{2}c_{1, 1}.
  }
  \label{eq:Tpol_c11}
\end{equation}
Upon inserting equation~\eref{eq:c11_2_clique_voter}
into equation~\eref{eq:Tpol_c11}, we obtain
equation~\eref{eq:Tpol_sparse_2_clique_voter}.

\section*{References}
\bibliography{voter_model_with_2_communities}

\end{document}